\documentclass[linespread 2]{article}
\title{On the satiability of floating bodies  }
\author { Mohammad Abolhassani \\  Department of Physics,  University of Arak,
Sardasht, 38156,\\ Arak, Iran. e-mail: m-abolhassani@araku.ac.ir.}
\usepackage{graphicx}

\begin{document}


\maketitle
\begin{abstract}

The potential energy of a system in stable equilibrium has a minimum
value. This property is used to derive a formula that is useful in
determination of stability of a floating body. It is found that a
floating body is in stable equilibrium if its center of gravity has
a minimum height with respect to its related center of buoyancy.

 \end{abstract}


keywords: stability, floating body, center of buoyancy, center of gravity, potential energy 

\section{\label{sec:level1}Introduction}

Study of stability of floating bodies is a conventional subject in
fluid mechanics. Two forces act on a floating object; weight, that
acts on the center of gravity $G$, and buoyancy force that acts on
the center of buoyancy $B$ (the centroid of the displaced volume of
fluid)~\cite{Halliday}. The equilibrium of a body requires that
these two forces be equal and opposite and the joint line of these
two points has to be in vertical direction too.

Stability of a floating body is divided in two different types,
vertical and rotational. A floating body has vertical stability but
its rotational stability depends upon the positions of $G$ and $B$.
If $G$ is below $B$ the equilibrium is stable. But if $G$ is above
$B$ the equilibrium may or may not be stable.
 The usual method in specification of stability of
a floating body is finding the metacenter point and then comparing
its position with $G$. The equilibrium is stable if the metacenter
lies above $G$~\cite{Streeter,Giles}.

Another approach to study this subject is using the energy
principle. Seemingly this idea first has been suggested by Chr.
Huygens in more than three centuries ago. P. Erd\"{o}s et al have
used this approach in solving some problems~\cite{Erdo} but they
have made some nonessential assumptions. In this paper stability of
floating bodies is studied by this method without restrictions
assumed in Ref.~\cite{Erdo}.

Consider a system consisting of a liquid and a floating object. If
the floating object is in stable equilibrium, the gravitational
potential energy of the system must be a minimum. Therefore, when
the object configuration is changed, the center of gravities of the
liquid and of the object will change; then the center of gravity of
the system raises and consequently the gravitational potential
energy of the system increases.

Inversely, if by changing  the  object configuration the
gravitational potential energy of the system increases, the object
is in stable equilibrium.

\section{Theoretical approach}

Consider a conservative system, with independent coordinates $q_1,
q_2, ....,q_n$, that is held in a  configuration and then released.
It generally  changes and gets a new configuration. But, if in
special cases, it does not change it will be in equilibrium and its
potential energy, $U$, is extremum with respect to all independent
coordinates of the system at its equilibrium configuration, $q_{01},
q_{02},....,q_{0n}$ ~\cite{Gold}, that is

\begin{equation}\label{1}
    \left(\frac{\partial U}{\partial q_{k}}\right)_{0}=0;\
    k=0,1,2,...,n.
\end{equation}

Figure 1 shows an object floating in a liquid. Masses of object and
liquid are $m$ and $M$ respectively. $\textbf{R}_{m}$,
$\textbf{R}_{M}$ and $\textbf{R}_{m+M}$ are the position vectors of
center of gravity of object, liquid and system (object + liquid)
respectively with respect to an arbitrary origin $O$.
 The gravitational potential energy of the system is given by

\begin{equation}\label{2}
    U=-(m+M)\textbf{g}.\textbf{R}_{m+M},
\end{equation}
where $\textbf{g}$ is acceleration of gravity.  $\textbf{R}_{m+M}$
can be written in term of $\textbf{R}_{m}$ and  $\textbf{R}_{M}$
~\cite{Fowles}

\begin{figure}[!htb]
\begin{center}
 \includegraphics[width=0.4\linewidth,angle=0,clip=true]{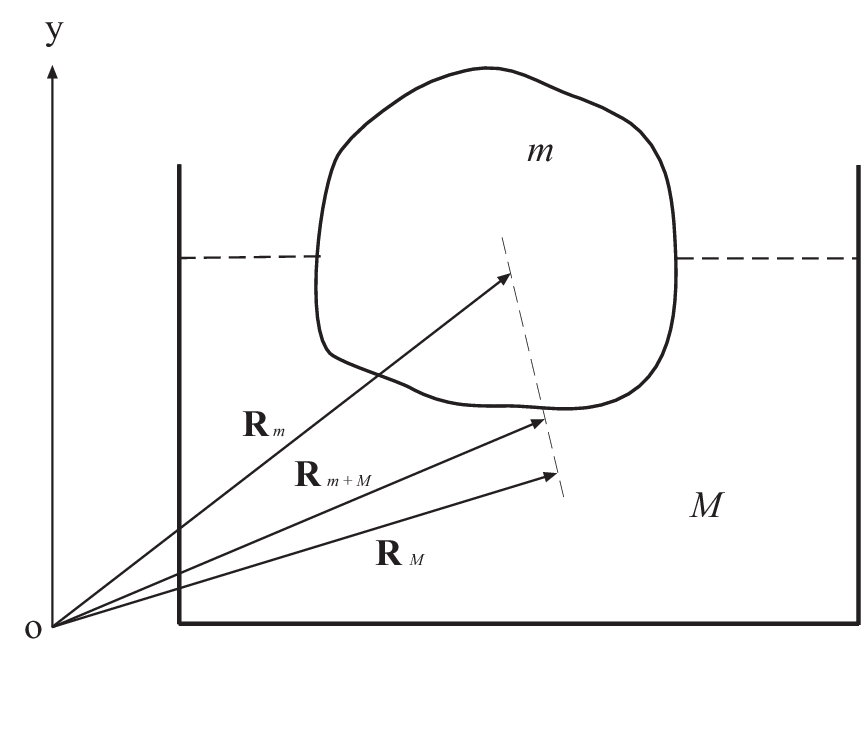}
\end{center}
\vspace*{-0pt} \caption{A floating object of mass $m$   is partially
immersed in a liquid of mass $M$. $\textbf R_{m}$, $\textbf R_{M}$
and $\textbf R_{m+M}$ are the position vectors of the center of
gravities of the object, liquid and system respectively.}
\label{fig1}
\end{figure}

\begin{equation}\label{3}
    (m+M)\textbf{R}_{m+M}=m\textbf{R}_{m}+M\textbf{R}_{M}.
\end{equation}
Thus
\begin{equation}\label{4}
    U=-(m\textbf{g}.{\bf R}_{m}+M{\bf g}.{\bf R}_{M}).
\end{equation}
Assume $y$ axis be in upward vertical direction, therefore

\begin{equation}\label{5}
    U=mgy_{m}+Mgy_{M},
\end{equation}
where $y_{m}$  and $y_{M}$  are $y$  coordinates of $\textbf{R}_{m}$
and  $\textbf{R}_{M}$ respectively.

 When the object is partially
immersed in the liquid, some part of the liquid is displaced by the
object. We name mass of the displaced liquid by $M'$  and the $y$
coordinate of its center of gravity, which is buoyancy center, by
$y_{M'}$. Figure 2 shows a hole in the liquid formerly occupied by
the immersed part of the object. If this hole to be filled by the
same type of liquid, so the mass of resulted liquid is "$M+M'$" and
 $y$  coordinate of its center of gravity  is $y_{M+M'}$. But,
according to the statement quoted before Eq. (3),

\begin{equation}\label{6}
    (M+M')y_{M+M'}=My_{M}+M'y_{M'}.
\end{equation}
\begin{figure}[!htb]
\begin{center}
 \includegraphics[width=0.4\linewidth,angle=0,clip=true]{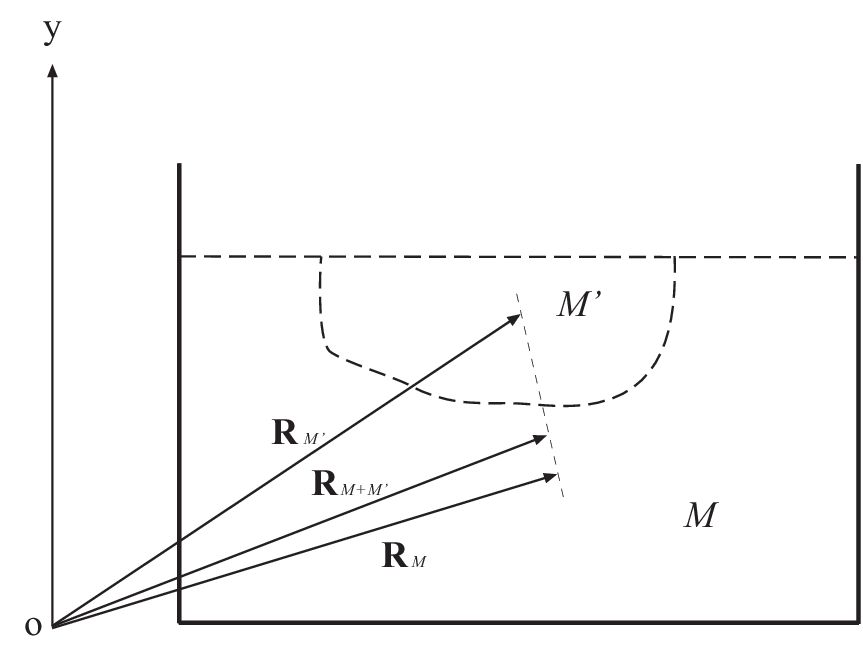}
\end{center}
\vspace*{-0pt} \caption{Initial liquid and the hole in the liquid
formerly occupied by the immersed part of the object. The hole is
filled by the same type of liquid. The mass of added liquid is $M'$.
$\textbf{R}_{M}$, $\textbf{R}_{M'}$ and $\textbf{R}_{M+M'}$ are
center of gravities of initial, added and resulted liquids
respectively.} \label{fig2}
\end{figure}
By eliminating $y_{M}$  between Eqs. (5) and (6), we have

\begin{equation}\label{7}
    U=\left[my_{m}+(M+M')y_{M+M'}-M'y_{M'}\right]g.
\end{equation}

This object has six degrees of freedom and then is described by six
independent coordinates; for example, three coordinates of its
center of mass (or gravity) and three angles that describe its
orientation. It is evident that the potential energy of the system
is independent of the object rotation around any vertical axis, and
the two coordinates that describe its center of mass in horizontal
plane. An equilibrium condition is

\begin{equation}\label{8}
    \frac{\partial U}{\partial y_{m}}=0.
\end{equation}
Now I calculate $\frac{\partial U}{\partial y_{m}}$ explicitly.
\begin{equation}\label{9}
    \frac{\partial U}{\partial y_{m}}=m+\frac{\partial M'}{\partial y_m}y_{M+M'}+(M+M')\frac{\partial y_{M+M'}}{\partial
    y_m}-y_{M'}\frac{\partial M' }{\partial y_m}-M'\frac{\partial y_{M'}}{\partial
    y_m}.
    \end{equation}
Figure 3 shows two configurations of the system. Assume areas of
liquid surface and cross section of object in liquid surface level
are $S$ and $s$ respectively. In the first case liquid surface is at
$y_0$. In the second case object has been raised by $\Delta y_m$ in
vertical direction and the liquid surface level has lowered by
$|\Delta y_0|$ (dashed shape). By inspection in this Figure we find
\begin{equation}\label{10}
\Delta M'=-\rho s(\Delta y_m-\Delta y_0),
\end{equation}
where $\rho$ is liquid mass density, and

\begin{figure}[!htb]
\begin{center}
 \includegraphics[width=0.4\linewidth,angle=0,clip=true]{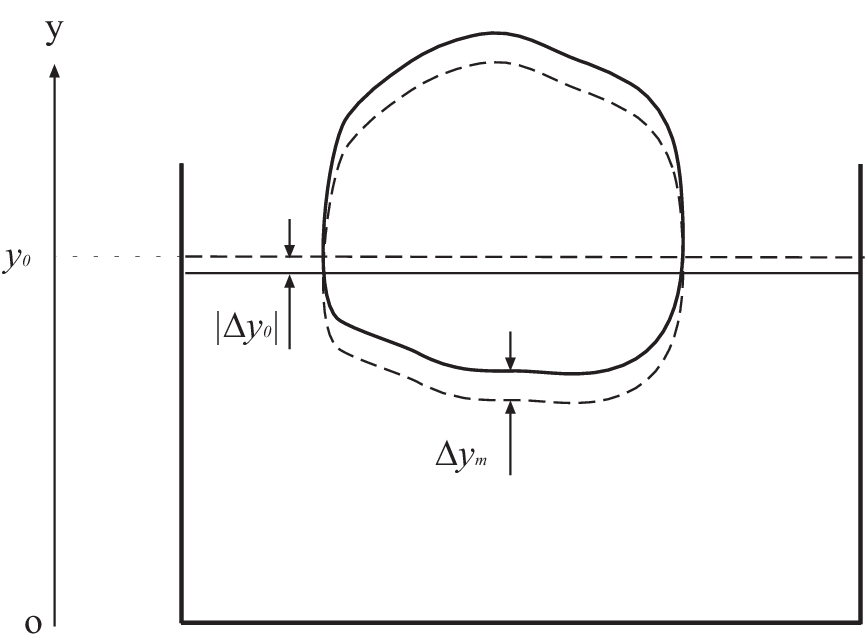}
\end{center}
\vspace*{-0pt} \caption{Two configurations of the system; before
raising (dashed), where the liquid surface level is at $y_0$, and
after raising the object by $\Delta y_m$ in vertical direction
(solid). In the latter case the liquid surface level  lower by
$|\Delta y_0|$.} \label{fig3}
\end{figure}

\begin{figure}[!htb]
\begin{center}
 \includegraphics[width=0.4\linewidth,angle=0,clip=true]{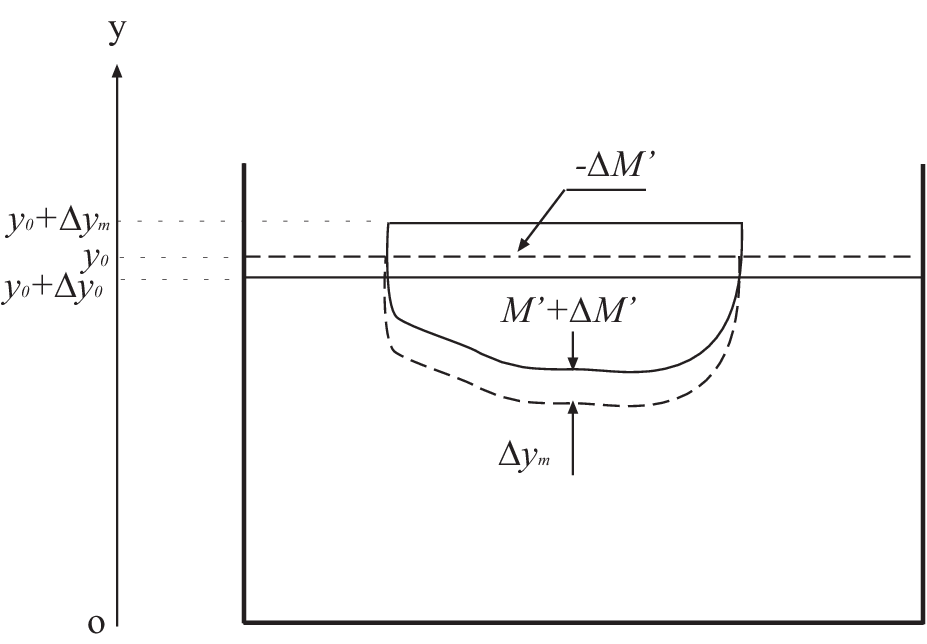}
\end{center}
\vspace*{-0pt} \caption{Added liquid in two configurations, before
 (dashed) and after (solid) raising the object.} \label{fig4}
\end{figure}
\begin{equation}\label{11}
\Delta y_0=-\frac{s}{S-s}\Delta y_m,
\end{equation}
therefore
\begin{equation}\label{12}
    \frac{\partial M' }{\partial y_m}=-\rho \frac{Ss}{S-s}.
\end{equation}

By raising the object, both $M'$ and $y_{M'}$ will change. This is
equivalent to raising $M'$ by $\Delta y_m$ and then shearing its top
by $\Delta y_m-\Delta y_0$, Figure 4. Relation between these two
parts according to center of mass theorem is

\begin{equation}\label{13}
    M'(y_{M'}+\Delta y_m)=(M'+\Delta M')(y_{M'}+\Delta y_{M'})-\Delta M'(y_0 +\frac{1}{2}\Delta
    y_m).
\end{equation}
By ignoring second order differential terms and introducing
$\frac{\partial {M'}}{\partial y_{m}}$ given by (12) into (13), it
follows that
\begin{equation}\label{14}
    \frac{\partial y_{M'} }{\partial y_m}=1+\frac{\rho}{M'}
    \frac{Ss}{S-s}(y_{M'}-y_0).
\end{equation}

Now by referring to Figure 5, that shows resulted liquid in two
cases, and repeating the procedure in writing Eq. (13) we have
\begin{figure}[!htb]
\begin{center}
 \includegraphics[width=0.4\linewidth,angle=0,clip=true]{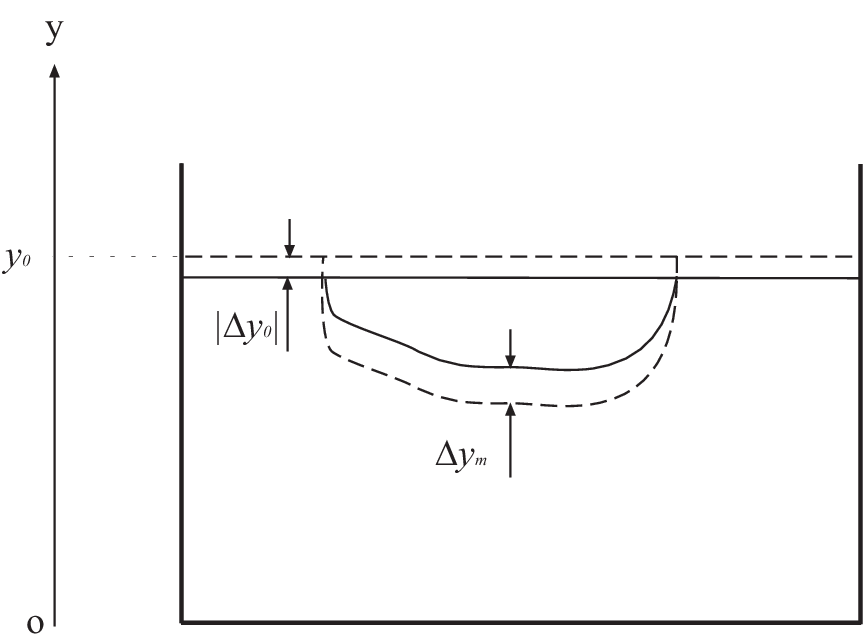}
\end{center}
\vspace*{-0pt} \caption{The resulted liquid in two cases, before
  (dashed) and after (solid) raising the object.} \label{fig5}
\end{figure}

\begin{equation}\label{15}
    (M+M')y_{M+M'}=(M+M'+\Delta M')(y_{M+M'}+\Delta y_{M+M'})-\Delta M'(y_0 +\frac{1}{2} \Delta
    y_0).
\end{equation}
By ignoring second order differential terms and using Eq. (12)  we
have

\begin{equation}\label{16}
    \frac{\partial y_{M+M'} }{\partial y_m}=\frac{\rho}{M+M'}
    \frac{Ss}{S-s}(y_{M+M'}-y_0).
\end{equation}
Substituting from Eqs. (12), (14) and (16) into (9) it reduces to

\begin{equation}\label{17}
    \frac{\partial U }{\partial y_m}=m-M'.
\end{equation}
Combination of (8) and (17) gives
\begin{equation}\label{18}
   M'=m.
\end{equation}
This is Archimedes' principle, which requires that in equilibrium
the mass of the displaced liquid $M'$ is equal to the mass of the
floating object $m$, independent of orientation of the object.
Therefore in vertical equilibrium $M'$ and also $y_{M+M'}$ are
constants. It is easy to derive this relation by another method but
this method is instructive and may be useful in studying dynamics of
floating bodies. Furthermore,
\begin{equation}\label{19}
    \frac{\partial^{2} U }{\partial y_m^{2}}=-\frac{\partial M' }{\partial
    y_m}=\rho \frac{Ss}{S-s}
\end{equation}
is always positive. Thus, equilibrium is always stable against
vertical displacements. By applying this result to Eq. (7) it
becomes

\begin{equation}\label{20}
    U=[(M+m)y_{M+M'}+m(y_{m}-y_{M'})]g.
\end{equation}
The first term on the right side of Eq. (20) is a constant.
Therefore gravitational potential energy of the system is a function
of $\Delta y=y_{m}-y_{M'}$.
 In case of $\Delta y$ is an extremum, the object is in equilibrium.
 When  $\Delta y$ is minimum
(maximum), $U$ is minimum (maximum) and therefore object is in
stable (unstable) equilibrium. Similar to this result ($BG$ instead
of $\Delta y$) has been quoted in Ref. ~\cite{Delbourgo} but has not
been derived by present method.

Another equilibrium condition is extremizing the potential energy of
system with respect to rotation of object around any axes that is
parallel to the liquid surface. This is equivalent to two equations.
The following example clarifies how this method can be used in
specification of stability of a floating body.

\section{example}

 A long uniform bar of square cross section each sides $a$, floats in liquid with its
longitudinal axis parallel to the liquid surface. Ratio of the
specific masses of the solid ($\rho'$)and liquid ($\rho$) is denoted
by $r=\frac{\rho'}{\rho}$. Bar has vertical stability and we wish to
determine its configurations in equilibrium condition. For this
purpose $\Delta y$ must be determined as a function of the variable
$\theta$, which is the angle between the symmetry plane of the bar
shown in Figure 6 and the vertical plane passing through the
longitudinal axis of the bar.  $\theta$ is restricted to the range
$[0, \pi/4]$, because increasing  $\theta$ over $\pi/4$ the bar
practically gets one of the old configurations.

An interchange $r\leftrightarrow (1-r)$ corresponds to
$v_{1}\leftrightarrow v_{2}$, where $v_{1} $ denotes  the immersed
volume and $v_{2}$ the exposed volume. Hence to each equilibrium
configuration there is a dual one where $r\rightarrow (1-r)$ and the
body is rotated through $180^{\circ}$. In fact it is proved that
duality preserves stability ~\cite{Erdo, Delbourgo}. Therefore I
shall restrict the discussion to $r\leq \frac{1}{2}$. For difference
in geometries, calculating of $\Delta y$ is made in two stages.

\subsection{Configuration with two corners immersed}
Figure 6a shows the bar with its two corners immersed. In this case
we have
\begin{equation}\label{21}
     y_m=a\left(\frac{1}{2}-r\right)\cos\theta
\end{equation}
and

\begin{equation}\label{22}
    y_{M'}=-a\left(\frac{1}{2}r\cos\theta+\frac{1}{24}\sin\theta\tan\theta\right),
\end{equation}
then

\begin{equation}\label{23}
    \Delta y=a\left[\frac{1}{2}(1-r)\cos\theta+\frac{1}{24r}\sin\theta\tan\theta\right].
\end{equation}
By taking the derivative of $\Delta y$ with respect to $\theta$ and
then equating to zero we find

\begin{figure}[!htb]
\begin{center}
 \includegraphics[width=0.8\linewidth,angle=0,clip=true]{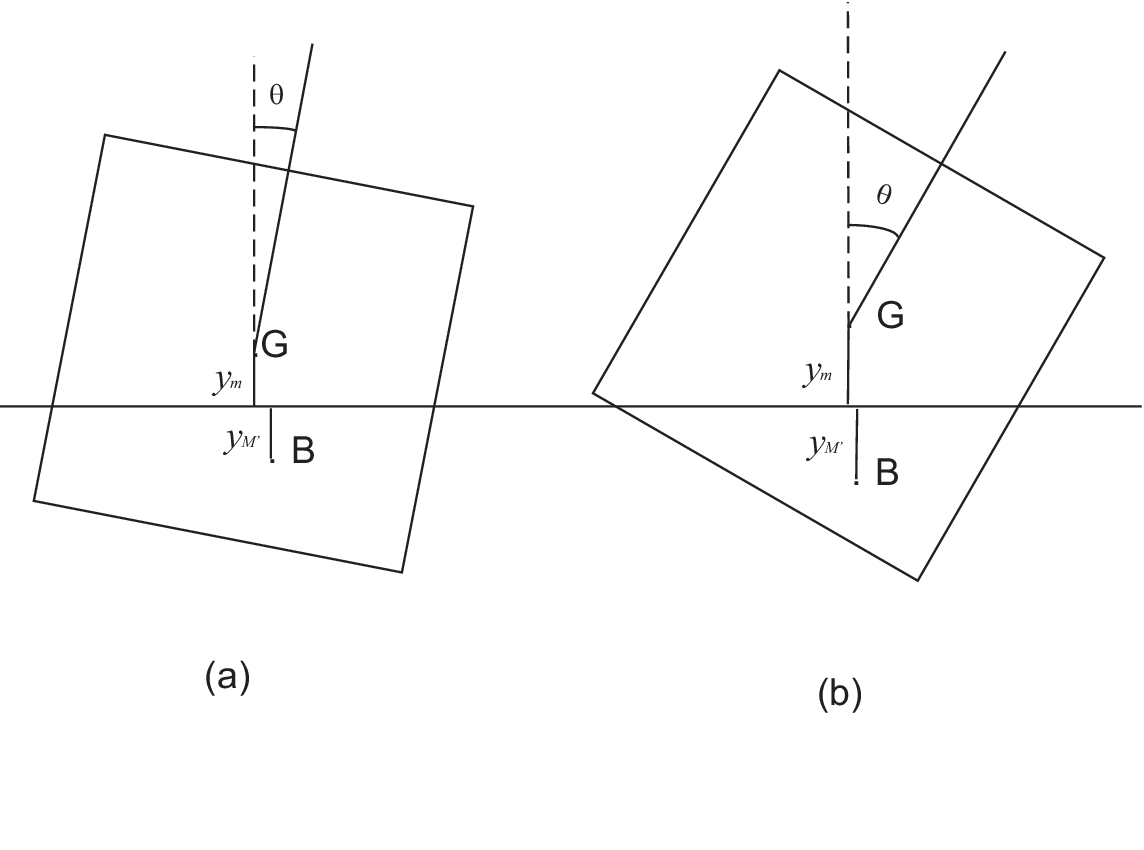}
\end{center}
\vspace*{-0pt} \caption{Vertical cut across  a bar floating in the
liquid; (a) two corners are immersed, (b) one corner is immersed.}
\label{fig6}
\end{figure}
\begin{equation}\label{24}
    \sin\theta[\tan^{2}\theta-12r(1-r)+2]=0.
\end{equation}
There are two solutions (a) and (b) of (24):

(a)

\begin{equation}\label{25}
    \sin\theta=0.
\end{equation}
Since

\begin{equation}\label{26}
\frac{\partial^{2}\Delta
y}{\partial\theta^{2}}|_{\theta=0}=\frac{a}{24r}[12r(r-1)+2],
\end{equation}
this equilibrium position is stable as long as
$r<\frac{3-\sqrt{3}}{6}$, but is unstable for
$\frac{3-\sqrt{3}}{6}<r\leq\frac{1}{2}$.

 (b) If $\theta\neq0$, Eq. (24) leads to

\begin{equation}\label{27}
    \tan\theta=\sqrt{12r(1-r)-2}.
\end{equation}
This solution has a real answer if
$\frac{3-\sqrt{3}}{6}<r<\frac{1}{2}$. On the other hand, because
both corners are immersed by assumption, there is an upper limit to
$\theta$, given by

\begin{equation}\label{28}
\tan \theta <2r.
\end{equation}
This condition with Eq. (27) leads to $r<\frac{1}{4}$. This means
that existence of this extremum requires
\begin{equation}\label{29}
0.21132\approx\frac{3-\sqrt{3}}{6}<r<\frac{1}{4}.
\end{equation}
Since
\begin{equation}\label{30}
\frac{\partial^{2}\Delta
y}{\partial\theta^{2}}|_{\tan\theta=\sqrt{12r(1-r)-2}
}=\frac{a}{12r}[12r(1-r)-2]\sqrt{12r(1-r)-1}
\end{equation}
is positive in the range of $r$ where second solution exists, this
equilibrium position (if exists) always is stable.
\subsection{Configuration with one corner immersed}

In this case we have, Figure 6b,
\begin{equation}\label{31}
     y_m=a\left[\frac{1}{2}(\cos\theta+\sin\theta)-\sqrt{r}\sqrt{\sin2\theta}\right]
\end{equation}
and
\begin{equation}\label{32}
    y_{M'}=-\frac{\sqrt{r}}{3}a\sqrt{\sin2\theta},
\end{equation}
then

\begin{equation}\label{33}
    \Delta y=\frac{1}{6}a\left(3\cos\theta+3\sin\theta-4\sqrt{r}\sqrt{\sin2\theta}\right).
\end{equation}
It must be noted that in this case the angle $\theta$ is restricted
to
\begin{equation}\label{34}
\tan \theta >2r.
\end{equation}
 It is more convenient for the following to express $\Delta y$ in
terms of the angle $\beta$, defined by
\begin{equation}\label{35}
    \beta=\pi/4-\theta.
\end{equation}
We have then

\begin{equation}\label{36}
\Delta
y=\frac{1}{6}a\left(3\sqrt{2}\cos\beta-4\sqrt{r}\sqrt{\cos2\beta}\right).
\end{equation}
 The equilibrium condition yields
\begin{equation}\label{37}
  \sin\beta\left(8\sqrt{r}\frac{\cos\beta}{\sqrt{\cos2\beta}}-3\sqrt{2}\right)=0.
\end{equation}
The two solutions, (c) and (d) are

(c)
\begin{equation}\label{38}
   \sin\beta=0, \     i.e., \theta=\frac{\pi}{4}.
\end{equation}
Since

\begin{equation}\label{39}
\frac{\partial^{2}\Delta
y}{\partial\beta^{2}}|_{\beta=0}=\frac{1}{6}a(8\sqrt{r}-3\sqrt{2}),
\end{equation}
this equilibrium position is stable as long as
$\frac{9}{32}<r<\frac{1}{2}$, meanwhile in this range condition (34)
is satisfied.

(d) For $\theta \neq \pi/4$
\begin{equation}\label{40}
    \cos2\beta=\frac{16r}{9-16r}.
\end{equation}
The conditions (34) and $|\cos2\beta|\leq1$ restrict the range of
$r$ to $\frac{1}{4}<r<\frac{9}{32}$. For this solution

\begin{equation}\label{41}
\frac{\partial^{2}\Delta
y}{\partial\beta^{2}}|_{\cos2\beta=\frac{16r}{9-16r}}=\frac{1}{6}a
\left(\frac{512r^{2}-432r+81}{16r\sqrt{9-16r}}\right),
\end{equation}
which is positive in $\frac{1}{4}<r<\frac{9}{32}$, hence this
equilibrium position is stable (if exists). For more details in
description of this and few other examples you can refer to
Ref.~\cite{Erdo}

\section{Conclusion}
Stability of a floating body was inspected by energy principle. This
formulation enables us to solve such problems without refereing to
the concept of the metacenter.  Archimedes' principle was derived
too by considering the energy principle. If the vertical distance
between center of gravity of the body and the center of buoyancy is
calculated versus a set of suitable coordinates, it is
straightforward to find any information about equilibrium positions
of the body.

\bibliography{apssamp}

\begin{thebibliography}{99}
%

\bibitem{Halliday} D. Halliday and R. Resnick. Physics.
 John Wiley \& Sons. 1978. chap. 17.


\bibitem{Streeter} V. L. Streeter, E. B. Wylie and K. W. Bedford. Fluid Mechanics.
 McGraw-Hill, New York. 1998. pp. 68-70.



\bibitem{Giles} R. V. Giles, J. B. Evett and C. Liu. Fluid Mechanics and
 Hydraulics. McGraw-Hill, Singapore. 1994. p. 58.


\bibitem {Erdo} P. Erd\"{o}s, G. Schibler and R.C. Herndon. Am. J. Phys. {\bf60}, 335 (1992).



\bibitem{Gold} H. Goldstein. Classical Mechanics.
 Addison-Wesley, Philippines. 1980. chap. 6.



\bibitem{Fowles} G. R. Fowles. Analytical Mechanics. CBS College Publishing, Japan, 1986. p. 191.

\bibitem {Delbourgo} R. Delbourgo. Am. J. Phys. {\bf55}, 799 (1987).

\end{thebibliography}

\end{document}